# Discovery of seven volcanic outbursts on Io from an IRTF observation campaign 2016 to 2022

Christian D. Tate[1], Julie A. Rathbun[1,2], Alexander G. Hayes[1], John R. Spencer[3], Madeline Pettine[1]

1 Department of Astronomy, Cornell University, Space Science Building, Ithaca, NY, USA
2 Planetary Science Institute, USA
3 Southwest Research Institute Boulder, Colorado, USA

## Abstract

This study analyzes near-infrared measurements of Io, Jupiter's moon, observed over 170 nights from 2016 to early 2022 using the NASA Infrared Telescope Facility (IRTF). During this period, seven new volcanic outbursts - the most energetic volcanic events on Io - were discovered and characterized, increasing the total number of observed outburst events from 18 to 25. We also present simplified criteria for the thermal detection of an outburst, requiring it to be both confined to a specific location of Io and above a threshold intensity in the Lp-band (3.8 μm).

Our measurements use 2-5 μm photometry in eclipse, Jupiter occultation, and reflected sunlight. In addition to extending the observational dataset of Io's dynamic activity, these data provide insights into the temporal and spatial distribution of outbursts on Io. Notably, all seven outbursts were detected in Io's trailing hemisphere. These include Pillan Patera and a newly discovered repeating outburst location at Acala Fluctus. We add these events to the rare category of recurring outbursts, before which Tvashtar was the only known example. We observed that another outburst at UP 254W decreased in Lp-band intensity by a factor of two in 4.5 hours. In August 2021, Io exhibited high volcanic activity when two powerful outbursts rapidly appeared, propagating East. Our findings underscore IRTF's ongoing contributions to the study of Io.

## 1 Introduction

Io attracts significant attention from both ground-based and space-based observatories, as well as planetary spacecraft. This is not just because Io is the most volcanically active body in the solar system but also because some of the most extreme forms of volcanism routinely appear on its surface. Various categories of volcanic activity have been observed on Io, including intra-patera, flow-dominated, and explosion-dominated eruptions (Williams and Howell, 2007; Veeder et al., 2012). Volcanic outbursts are the brightest events with an eruptive power that is more than an order of magnitude higher than Io's typical volcanic hot spot. Despite their immense peak brightness, outbursts are short-lived, and their time-averaged contribution to Io's total radiation is only 1-2% (Veeder et al., 2012). Prior to this work, there were 18 outburst events on record (Veeder et al., 2012, Cantrall et al., 2018). Spencer and Schneider (1996) estimated a frequency of about one occurrence per month, which they calculated from the combined detection rate of several early campaigns from 1978 to 1995. During these observations, there was about a 1.8% chance of observing an outburst on any given night. These early estimates have held up surprisingly well in the following decades. Once an outburst is observed, it generally fades into Io's background emission within several days (de Kleer and de Pater (2016a), Davies et al., 2001).

Despite not contributing significantly to Io's time-averaged total emission, outbursts represent the most energetic events observed on Io and dominate its 2-5 µm spectrum. They reach estimated effective temperatures of 1000-2000 K, with the highest temperatures (> 1500 K) requiring ultramafic lava compositions (Davies et al., 2001; Keszthelyi et al., 2007; Keszthelyi and Suer 2023). Although most of Io's heat loss is through the more numbers but individually less energetic volcanic hot spots (McEwen et al., 1998; Lopes et al., 2005), understanding the mechanisms responsible for generating high-power and high-temperature outbursts is critical to a holistic understanding of Io's volcanic system.

Herein, we describe the results of an observation campaign conducted between June 2016 and January 2022 designed to monitor Io during NASA's Juno mission. These observations are part of a long-term Io monitoring campaign conducted using NASA's Infrared Telescope Facility (IRTF) on Mauna Kea's sacred and arid summit in Hawaii. We describe our campaign methodology and discuss the seven new outburst events, including their constrained location, temperature, and brightness.

## 1.1 Definitions of outbursts and mini-outbursts

Based on disk-integrated observations, Sinton et al., (1983) proposed an early definition for outbursts in terms of Io's effective geometric albedo. Spencer and Schneider (1996) took this concept one step further and defined an outburst as a doubling of Io's 5-µm brightness. This became the canonical definition, well suited for the infrared astronomy techniques available on the eve of the Galileo era. Adopting a 5 µm standard was natural because of its proximity to the peak wavelength of blackbody emission temperatures of the first observed outbursts.

Despite the simplicity of this definition, the doubling of Io's 5-µm brightness is ambiguous in several regards. Firstly, the specific value that should be doubled to qualify an outburst is not explicitly quantified. Secondly, this definition is agnostic regarding what precisely constitutes an event in spatial dimensions. An outburst event could be a single volcanic eruption or multiple contemporaneous eruptions scattered over the observable hemisphere. There is also an observational difficulty of obtaining high signal-to-noise ratio measurements at 5 µm due to Earth's atmospheric opacity. A more favorable wavelength is the 3.8 µm Lprime-band. Io's near-infrared albedo is also difficult to quantify due to geomorphological changes. Io has a complicated and poorly constrained phase function, including a significant opposition surge. At this point, these effects are better understood at 3.8 µm than 5 µm.

The category of mini-outbursts was coined by de Kleer and de Pater (2016b) and, unlike the original outburst definition, is based on the brightness of a single volcano and not on the total disk-integrated brightness of Io. This population includes transient eruptions that surpass a peak brightness of 30 GW/sr/µm in Lprime-band, which roughly is equivalent to the combined intensity of all visible hot spots at a given time and less than the typical outburst by a factor of about five.

A revised definition of outburst and mini-outburst is necessary due to advancements in high-resolution and frequent imaging of Io's volcanic activity over the last twenty years. Our proposed definition, like that of de Kleer and de Pater (2016b), is based on Lp-band (3.8 μm) intensity. We define an outburst as a hot spot with a 3.8 μm intensity greater than 150 GW/sr/μm. This value refers to the intensity of the individual hot spot and not the disk-integrated intensity. The threshold between outbursts and mini-outbursts is also somewhat arbitrary, although it has an empirical justification. This threshold at 150 GW/sr/μm encompasses most previously reported outbursts (Spencer and Schneider, 1996; Veeder et al., 2012; Cantrall et al., 2018). We emphasize that this is a new and simplified definition intended to standardize the characterization of Io's bright transient eruptions. Therefore, we define a large transient eruption with two criteria: the thermal detection of an outburst or mini-outburst on Io requires evidence that the eruption is bright and confined to a small region.

1. An outburst has a *large* thermal intensity, $I_{3.8μm}$ > 150 GW/sr/μm
2. An outburst intensity must come from a *small* volcanic source.

The second constraint requires that the event is reasonably confined to a single source or small region on Io. Because the spatial resolution quality varies significantly between observation techniques and generations of instrumentation, the upper limit for the localization uncertainty depends on the method used. An outburst's uncertainty can be as high as 20° in the longitude and latitude directions (or about 640 km for events at high latitudes) for IRTF's resolved imaging. For occultation-reappearance detections, however, the constraint is 5-10° but only in the direction perpendicular to the limb of the occluding body (Jupiter). The upper limit for a one-dimensional constraint should be better than about 10° (or about 320 km). In this definition, a disk-integrated intensity measurement alone cannot fully qualify an event as an outburst. There must be evidence that the high intensity comes from a reasonably small region on Io.

## 2 Data Collection

Between June 2016 and August 2022, we observed Io on about 170 nights. Table 1 lists the number of observations during each proposal cycle. While our observation techniques are based on those by Spencer et al., (1990), the IRTF instrumentation and processing chain has significantly changed over the intervening 35 years. Accordingly, we explain our observations and data analysis below. We focus on the novel portions of our processing chain, such as recovering Io's endogenic thermal emission from sunlight imaging. A detailed review of the entire processing chain, including the more traditional portions, is provided in the Supplementary Online Material for this paper.

Our IRTF campaign employs three observation modes or strategies to monitor Io's volcanic activity: first imaging Io while it is eclipsed by Jupiter's shadow; second obtaining occultation light curves from movies while Io passes behind Jupiter's limb and is simultaneously eclipsed from the Sun; and third capturing images while Io is fully illuminated by sunlight. The first mode serves as our primary means of measuring Io's full-disk volcanic activity, while the second mode provides valuable spatial information about the hot spots distributed across Io's disk. Modes one and two yield measurements of only the sub-Jovian hemisphere of Io. The third mode, sunlit imaging, observes all longitudes of Io. However, this mode is a highly challenging method for measuring volcanic activity with IRTF. Despite this, IRTF sunlit imaging of Io is adequate for detecting the brightest volcanic activity, such as volcanic outbursts (Spencer and Schneider 1996). In fact, this campaign detected outbursts in each of the three observation modes.

Each type of observation utilizes the guiding camera of either the SpeX or iShell instruments to obtain photometric images in a subset of the CH4l-band, contK-band, Lprime-band, and M-band wavelengths (Tokunaga and Vacca, 2005). The properties of these filters, and the images obtained in them, are given in Table 2. The images of Io and Jupiter (Figure 1) appear as white-black pairs because we take the image difference between two telescope positions that are 10 arcseconds apart causing Io and Jupiter to be in the field of view in both the image and sky frames. This method helps control the detector bias offset, the telescope's stray light, and other second-order effects essential for infrared photometry. We use this method mostly in eclipse mode. To calibrate the differences in the telescope, instrument, and sky from night to night, we image a standard star on each observing night, regardless of the modes used. These standard stars have known absolute magnitudes in at least the K band, and this information calibrates the measurement of Io's brightness.

The first two observation modes (eclipse and occultation) necessitate precise scheduling to ensure that Io is in eclipse and visible from Earth during the observation run. Observation scheduling is also constrained by solar conjunctions, which preclude ground-based observation of Io for approximately 40 days during every synodic period of the Earth, Sun, and Jupiter system. This synodic period is 400 days, or roughly one year, one month, and five days. The shaded regions in Figures 2 and 3 illustrate when Io's position in the sky is too close to the Sun for photometric observations.

Eclipse observations and occultation curves primarily capture the sub-Jovian hemisphere of Io, centered at $0\pm10°$ longitude (refer to the vertical axis in Figure 3). Jupiter's shadow is a critical aspect of these measurements, as reflected sunlight would overwhelm the thermal signature of high-temperature hot spots in the 1.5-5 μm range. Moreover, Io's time and longitudinally varying albedo, phase curve, and emissivity are not accurately known enough to precisely subtract the Sun's irradiant contribution from Io's total brightness. It is important to note that these observations can only be used for the sub-Jovian hemisphere of Io.

In the third observing mode, we measure Io's brightness in full sunlight at quasi-random times throughout Io's orbit. This provides complete longitudinal coverage centered on Io's equator, with a sub-earth latitude range of $0\pm4°$ and phase angles from 0-12°. However, this strategy offers a much lower accuracy for measuring Io's thermal emission from the volcanic hot spots. Although this method accurately measures the total flux from Io, the reflected sunlight accounts for approximately 80% of the L-prime flux and 50% of the M flux when Io's hot spots are at their average output. In order to measure only the endogenic thermal emission, we estimate the reflected sunlight and subtract it from the total brightness.

# 3 Methods

## 3.1 Photometric Calibration

We make all intensity measurements of Io relative to one or more standard stars of known flux in at least the K-band (see Table s1 in SOM). The M-band magnitudes of many of these standard stars are not well constrained, however, and we chose type A0 standard stars to minimize the change in magnitude across the 1.6-5 µm range of this investigation. With this assumption, we use the star's known K or L-prime magnitude as a first order estimate for its M magnitude.

For the eclipse observations in Jupiter's shadow, Io's spectral intensity and its uncertainty in units of GW/sr/µm are,

$$I_{Io} = E_0\, 10^{-0.4\, M_{st}} \frac{f_{Io}}{f_{st}} C d^2, \tag{1}$$

where $E_0$ is the zero-magnitude irradiance given in Table 1; $M_{st}$ is the known absolute magnitude of the calibration star; $f_{Io}$ and $f_{st}$ are the detector counts for Io and the standard star measured from the IRTF images; C is the air-mass correction between Io and the standard star; and d is the distance between Earth and Io. Each CH4-band, contK-band, and Lp-band observation is calibrated with a standard star of known absolute magnitude.

We use the JPL Horizons tool to determine the distance from Earth to Io, $d$, the distance from the Sun to Io, $r$, Io's sub-observer longitude, latitude, and incidence, emission, and phase angles. These geometric parameters are necessary to convert Io's brightness measurements to absolute values of thermal emission from Io's surface, which in turn, we interpret as Io's hot spot activity.

The uncertainty of Io's intensity is the sum in quadrature of each uncertainty of each variable in equation 1. The approximate uncertainties for the Solar irradiance and air mass corrections are $\sigma_{E0}/E_0 \simeq 0.01$, $\sigma_C/C \simeq 0.01$, and $\sigma_{Mst} \simeq 0.1$. The brightness uncertainty includes an error contribution for the standard star's photometric magnitude uncertainty ($\sigma_{Mst}$) in each band because we often infer the star's photometric magnitude (Mst) from lower wavelength photometric bands. Due to this assumption, we prefer A0V stars because they have approximately the same magnitudes from 2-5 µm (Table s1 in SOM). When the L-band magnitude is known for a standard star, we use that magnitude for the L-prime and M bands, $M_{L\text{-prime}} \sim M_L$ and $M_M \sim M_L$. When only the K band magnitude is known, however, then that magnitude is used for all the longer wavelength bands. We quantize these errors by proportionally increasing $\sigma_{Mst}$ for the long wavelength standard star magnitudes that are most uncertain.

## 3.2 Occultation Light Curves

The occultation technique pioneered by Spencer et al., (1990) and used for more than thirty years (Rathbun et al., 2002, Rathbun and Spencer, 2010), provides a time-resolved distribution of brightness over the hemisphere of Io facing Earth. By frequently imaging Io in eclipse as it disappears or reappears behind Jupiter's limb, we measure a light curve for the fraction of Io visible at each instance of its occultation or apparition. We infer the brightness of each hot spot or cluster of hot spots from the height of the observed jump in the light curve. The right-hand side plots in Figure 6 show a sudden jump in brightness at the outburst location. The projection of Jupiter's limb on Io gives the approximate longitude of the hot spot and constrains the one-dimensional location on the axis of Io's relative motion to Jupiter. As seen in Figure 8, the projection of Jupiter's limb can tilt due to the phase of orbital inclinations between Earth, Jupiter, and Io.

| IRTF Semester | Eclipse Obs. (modes 1 and 2) | Sunlit Obs. (mode 3) | Solar Conj. | Outbursts Obs. in Eclipse (1 and 2) | Outbursts Obs. in Sunlight (3) |
|---|---|---|---|---|---|
| 2016A | – | 1 | – | – | – |
| 2016B | 2 | 7 | Aug – Oct | – | – |
| 2017A | 8 | 9 | – | – | – |
| 2017B | 4 | 9 | Sep – Nov | – | Jan 18 |
| 2018A | 8 | 7 | – | – | May 10 |
| 2018B | – | – | Oct – Dec | – | – |
| 2019A | 10 | 14 | – | May 8, June 25 | – |
| 2019B | 5 | 2 | Nov – Jan | – | – |
| 2020A | 8 | 9 | – | – | – |
| 2020B | 11 | 11 | Jan – Feb | Nov 15 | – |
| 2021A | 8 | 9 | – | – | – |
| 2021B | 7 | 9 | – | – | Aug 13, Aug 27 |
| 2022A | – | – | Feb – Mar | – | – |
| **Total** | **71** | **87** | – | **3** | **4** |

**Table 1**. The number of successful IRTF SpeX observations from 2016 to 2022. The totals are 71 eclipse and occultation observations (1.5 hours for modes 1 and 3) and 87 sunlit observations (0.5 hours each for mode 3) of the course of about 170 observation nights.

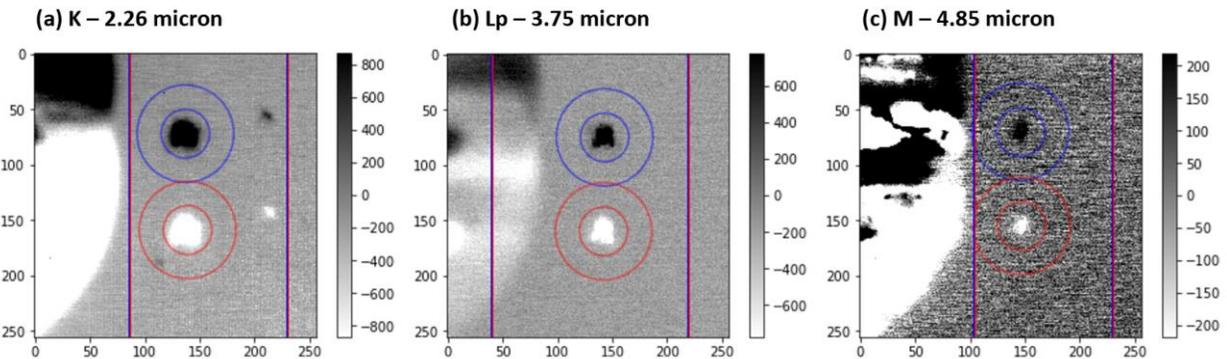

**Figure 1.** Images of Io and Jupiter from the IRTF SpeX camera during the outburst event of May 8th, 2019. Io is in eclipse. The images are the infrared photometric filters (a) narrowband ck 2.3 μm, (b) broadband L-prime 3.8 μm, and (c) broadband 4.5-5.1 μm M-band (see Table 2). The double-effect in the images is due to the subtraction of A and B frames when the telescope nods by 5 arcseconds in the north-south direction to remove the telescope's and atmosphere's first-order thermal effects. The frames are 30 arcseconds across. Notice the other Jovian moons visible in the contK-band image that disappear in the longer wavelength images. L-prime is used for the occultations both because of its sensitivity to thermal

anomalies on Io and because Jupiter itself is less bright due to its methane absorption. The signal-to-noise ratio is noticeably lower for the M-band images.

| IRTF filter bands | - | CH4_l | Kcont | Lprime | M |
|---|---|---|---|---|---|
| $\lambda$ wavelength | μm | 1.69 | 2.27 | 3.78 | 4.85 |
| $\lambda$ range | μm | 1.61 - 1.78 | 2.26 - 2.29 | 3.42 - 4.13 | 4.54 - 5.16 |
| Average intensity of Io in eclipse | GW/sr/μm | 7.6 | 8.4 | 34 | 102 |
| Average intensity of Io in reflected sunlight | GW/sr/μm | - | - | 732 | 446 |
| intensity of an outburst (definition) | GW/sr/μm | >50 | >50 | >150 | >400 |
| Mini-outburst (definition) | GW/sr/μm | >~10 | >~10 | >30 | >90 |
| Estimated average albedo of Io | - | - | - | 0.49 | 0.66 |
| Effective average albedo of Io | - | - | - | 0.54 | 0.87 |

**Table 2.** IRTF SpeX and iShell photometric filters were used during this observation campaign (Tokunaga and Vacca, 2005). For reference, we give our radiant calibration factors for each band and our median brightness values of Io in eclipse and sunlight.

## 3.3 Sunlit image processing

The sunlit observation mode has the advantage of resolving Io's full disk, which is valuable for constraining the location of hot spots when they are bright enough to distinguish from the disk itself. In contrast, the eclipse images of Io only show the hot spots without any clear boundary of Io's limb. Since IRTF does not have adaptive optics (AO), however, many sunlit images are distorted by atmospheric turbulence. We mitigate this effect by filtering out blurry images and aligning only the sharpest images. This shift-and-add technique is used to get a first-order image of Io. We then apply standard anti-blurring algorithms to sharpen the image.

The first and third columns of Figure 7 show these shift-and-add images with the white circle signifying the approximate limb of Io, which we guess by eye. The circles depicting the limb of Io are, on average, 9 pixels in diameter on the camera's focal plane. Since we know Io's sub-observer longitude, latitude, and relative inclination, we can map this image onto the globe of Io (as seen on the second and fourth columns of Figure 7) and localize the hot spots. The estimate of Io's limb on the shift-and-add images, however, is the greatest source of spatial uncertainty. We estimate the hot spot localization uncertainty by randomly offsetting the limb placement by two-dimensional Gaussian and calculating where the hot spot projects onto Io for each limb. The limb offset distribution has a standard deviation of two pixels in both the vertical and horizontal directions. From this estimate, the location uncertainties are between 12 to 15° depending on the hot spot's distance from the limb. This method nevertheless provides valuable estimates of the hot spots that are bright enough to outshine the reflected sunlit.

To preserve photometric accuracy after this processing, we rescale the shift-and-add image to the average of Io's total intensity, which is measured in the same photometric calibration processes used for the eclipse image mode. Although previous IRTF studies of Io did not quantitatively analyze Io's sunlit imaging, we aim to more rigorously constrain the thermal component of Io's total sunlit intensity. This is difficult because reflected sunlight overwhelms Io's thermal emission if not properly accounted for. Nevertheless, this process can reliably identify thermal Lp-band intensities greater than 150 GW/sr/µm needed to detect outbursts.

We analytically model the brightness of Io in sunlight as a combination of thermal emission and reflected sunlight. The total intensity of Io that IRTF measures is,

$$I_{total} = I_{thermal} + I_{sunlight}. \tag{2}$$

Because the reflected sunlight $I_{sunlight}$ is over an order of magnitude greater than Io's median thermal emission $I_{thermal}$ in the Lp-band (Table 2), great care must be taken to model $I_{sunlight}$ accurately enough to detect large scale volcanic eruptions. We estimate the intensity of Io's reflected sunlight as,

$$I_{sunlight}(\alpha, r, \lambda) = p\, C_p(\alpha) \frac{R^2}{r^2} E_{Solar}(\lambda), \tag{3}$$

where *p* is the longitudinally averaged geometric albedo, $C_p(\alpha)$ is the correction factor for the varying phase angles $\alpha$, *R* is the radius of Io, *r* is the distance from the Sun to Io in astronomical units, and $E_{Solar}$ is the solar irradiance at 1 astronomical unit. The product of p and $C_p(\alpha)$ is the fraction of sunlight reflected towards the observer relative to a perfectly reflective Lambertian disk the size of Io and at Io's distance from the Sun.

The greatest unknown in equation (4) is the product of the geometric albedo p and the phase angle correction $C_p(\alpha)$ and, consequently, previous studies have not provided robust estimates of Io's disk-integrated near-infrared photometric properties. Some studies circumvent this issue by removing reflected sunlight by subtracting brightness from smooth areas on Io's resolved disk (de Kleer 2016a). We cannot employ this method, however, because IRTF's optics do not sufficiently resolve Io's disk. A meaningful and much-needed contribution to the study of Io would be to rigorously measure Io's albedos and phase curves for the thermal wavelengths 1-5 µm.

Establishing a lower limit for the albedo, *p*, is challenging due to Io's significant ongoing activity. The M-band albedo, influenced by the low temperature of the background radiation, is more uncertain than the Lp-band albedo. Io also has varying spatial albedos that can evolve after large resurfacing events. These effects, however, are not well characterized, so we assume that Io's geometric albedo is longitudinally and temporally uniform within the uncertainty of our observations.

We set Io's physical geometric albedo, *p*, to be equal to the tenth percentile of the effective albedo derived from non-outburst observations in our sunlit dataset. We calculate an effective albedo for each sunlit observation as the ratio of Io's total intensity to a Lambertian disk (see Table 2).

Conversely, Io's average effective geometric albedo, p', given in Table 2, is the median of all sunlit observations in the dataset, including outbursts. These values are calculated from N=100 and N=19 data points for the Lp-band and M-band, respectively. Note that albedo values in Table 2 are for a phase angle of α=5°. The effective albedos are larger because they incorporate persistent background volcanic radiation. An alternative estimate can be derived from the difference between the median sunlit and eclipsed brightness values in Table 2. When these estimates are converted to effective albedos, they are similar to the values we assume in our analysis. From this data, we can find what fraction of Io's median total sunlit intensity is due to the average thermal activity on Io's surface. On average, approximately 5% and 31% of Io's total sunlit intensity in the Lp-band is due to thermal activity.

Since Io's scattering behavior has a significant phase angle dependence, we apply a secondary correction $C_p(\alpha)$ to account for Io's behavior across the range of $0.2 < \alpha < 13°$ (Figure 7). We parameterize $C_p(\alpha)$ with a polynomial fit to the non-outbursting geometric albedo values. Io reflects more than 20% more light at α=0 than α=5°. There appears to be a significant opposition surge for α < 0.5°, which $C_p(\alpha)$ under-estimates. Therefore, care must be taken when interpreting low-phase angle observations.

We apply this phase angle correction only to the Lp-band observations. Although we added M-band sunlit imaging to our data collection in 2021, the low number of data points prevents us from modeling Io's M-band phase curve similarly. M-band measurements are nevertheless valuable for their sensitivity to thermal activity. We do not measure Io sunlit contK-band brightness because the reflected sunlight overwhelms Io's typical hot spot brightness.

After accounting for the reflected sunlight, we estimate the thermal intensity of Io from the difference between the measured total and the calculated reflected sunlight intensity.

$$I_{thermal} = I_{total} - I_{sunlight}. \tag{4}$$

The hot spot intensity is then estimated as the difference between the thermal and time-median values,

$$I_{hs} = I_{thermal} - I_{thermal,\ median} \tag{5}$$

We assume the uncertainty on $I_{thermal}$ is not greater than the uncertainty of $I_{total}$, so we adopt the observational uncertainty of Io's total intensity to estimate its thermal uncertainty. While our measurements of Io's disk-integrated thermal intensity in sunlight have a relatively low signal-to-noise ratio, this method is sufficient to identify and localize volcanic outbursts.

$$\sigma_{hs} \simeq \sigma_{thermal} \simeq \sigma_{total} \tag{6}$$

Given the high uncertainties of these sunlit intensity measurements, however, we qualify an event as an outburst only if its lower limit of intensity exceeds the outburst threshold, $(I_{hs} - \sigma_{hs}) > I_{threshold}$.

## 3.4 Blackbody Temperature Fits

In eclipse observations, we generally measured the brightness of Io at three or four bands. After calculating Io's fully eclipsed brightness for each filter, we estimate the blackbody spectrum and the effective temperature of Io's hot spots. Since we work with disk-integrated brightness values, we assume that the outburst flux dominates Io's brightness in the 1-4 μm range. As a first-order correction to this assumption, we estimate the probable background brightness values by taking the mean intensities of our IRTF dataset when Io was not outbursting. These background values for each filter are in Table 2. We modify our direct measurements in two ways with these background values before fitting the blackbody model to the outburst spectra. First, we subtract them from the measured disk-integrated brightness values. Second, we add them to the measurement uncertainties.

The spectra of the outburst events often match a blackbody curve (de Kleer and de Pater, 2016a), especially for the Lp-band and the bands of shorter wavelengths (see Figure 6). By assuming an emissivity of ε = 1, the photometric intensities of several photometric bands can constrain the effective temperature and projected area of the hot spot(s) in a single-temperature fit (1-T). While this strategy truncates Io's wide temperature variability, it allows a valuable retrieval of an outburst's properties with 2-4 brightness measurements at different wavelengths. Given the nearly over-constrained degrees of freedom in these fits, their uncertainties are estimated from the image frame-by-frame noise of the brightness values.

$$I_{thermal}(\lambda) = A\epsilon \frac{2hc^2}{\lambda^5} \frac{1}{\exp \frac{hc}{\lambda k_B T} - 1} \tag{7}$$

Here, h is the Planck constant, c is the speed of light, $\lambda$ is the wavelength, $k_B$ is the Boltzmann constant, $A_{HS}$ is the hot spot's projected area, ε is the hot spot's emissivity assumed to be unity $\epsilon=1$, $T_{HS}$ the hot spot's effective temperature. The results of these fits are given in Table 3 and Figure 7.

## 4 Results

Our data support the discovery of seven outbursts, three were discovered in eclipse observations and four from observations of Io's sunlit hemisphere. We also observed several mini-outbursts (in March 2017, May 2021, and August 2021; Figure 7), all these are consistent with Loki Patera. Since Loki Patera is a special case (Rathbun and Spencer, 2006; Veeder et al., 2009; de Kleer and de Pater, 2016b, de Pater et al., 2017), we do not consider it in our analysis of outbursts.

### 4.1 Outbursts Observed in Eclipse and Occultation

Using our two-criterion definition described above, we identify three outbursts in the eclipse and occultation measurements of Io. Figure 2 shows ellipse brightness values in the three primary photometric bands (contK, Lp, and M). Outburst events manifest as large brightness values at 2-3 µm (e.g., contK-band) with elevated values at longer wavelengths that surpass the Lp-band threshold, 150 GW/sr/µm. Compared to the shorter wavelengths, the relative enhancement of Io's brightness in the 3-5 µm range is less dramatic.

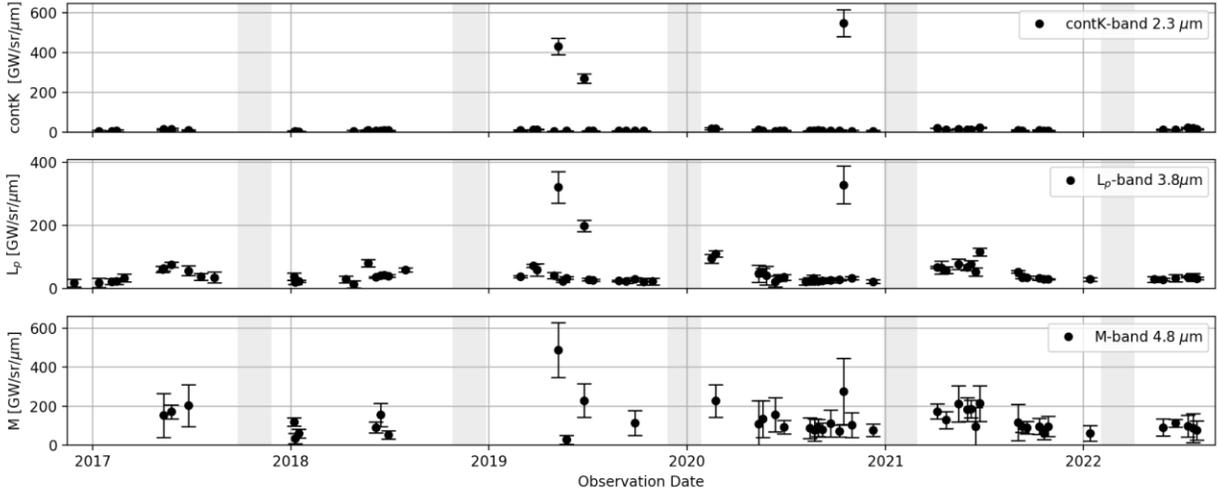

**Figure 2.** Time series of Io's eclipsed disk-integrated sub-Jovian intensity for 2016-2022 in three bands. The three outbursts in this plot stand out in the contK-band, which is near the peak of the spectral emissions for these volcanic events. The gray bands show the periods of time when Jupiter is in solar conjunction.

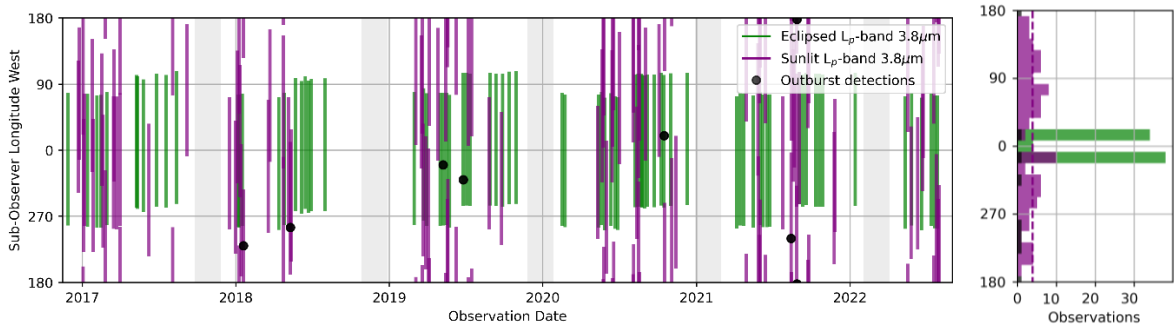

**Figure 3.** Time series of the sub-observer (e.g., central meridian) longitude +/- 90° for each of our eclipsed and sunlit observations. The black dots are the sub-observer longitudes at the time for each outburst event. The sub-figure on the right is the histogram of the observations divided into 15-degree sub-earth longitude bins. The gray bands show the periods of Jupiter's solar conjunction. The eclipsed observations are clustered around the sub-Jovian point at 0° West. The sunlit observations are scattered semi-evenly around Io with the dashed line showing the average of 3 sunlit observations per 15-degree bin (or approximately one per 5° longitude). Note that while we have more observations of the leading hemisphere (centered on 90 W longitude) outbursts favor the trailing hemisphere (centered on 270 W longitude).

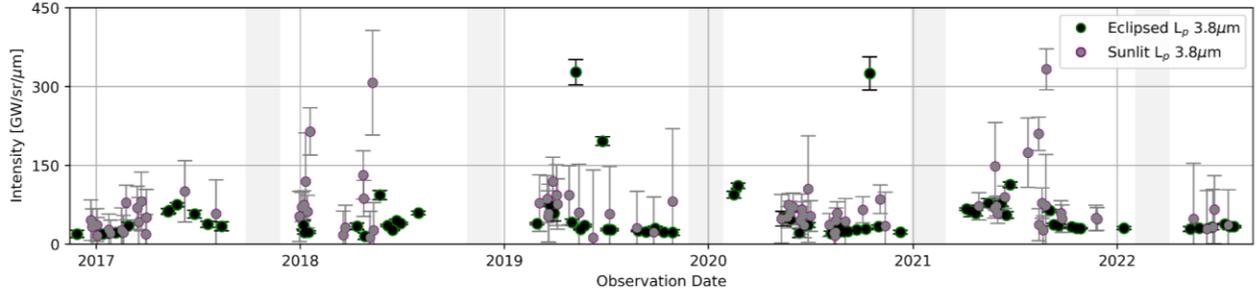

**Figure 4.** Time series of Io's total 3.8μm endogenic brightness values for both eclipse and sunlit observations (where the estimated contribution from reflected sunlight has been subtracted). The outburst events are the points above 150 GW/sr/μm. Despite the larger uncertainty in the sunlit observations, the events in May 2018 and August 2021 stand out. The gray bands show the periods of Jupiter's solar conjunction when Io is not visible from Earth.

| Universal Time | Event Name | Closest hot spot | Observing Mode | Latitude North | Longitude West | Phase angle | Emission angle | Emission corr. μ | Intensity CH4_l | Intensity contK | Intensity Lp | Intensity M | Intensity Lp corr. | Temp [K] | Area [km2] | Power [TW] |
|---|---|---|---|---|---|---|---|---|---|---|---|---|---|---|---|---|
| 2018-01-18 16:10 | 201801A | *Mulungu Patera?* | Sunlit imaging | 10 ± 14 | 220 ± 15 | 9.7 | 19 | 0.95 | – | – | 183 ± 45 | – | 194 ± 48 | – | – | – |
| 2018-05-10 10:20 | UP 254W | *UP 254W* | Sunlit imaging | -40 ± 14 | 255 ± 16 | 0.4 | 32 | 0.85 | – | – | 275 ± 99 | – | 331 ± 119 | – | – | – |
| 2019-05-08 14:25 | 201905A | Acala Fluctus | Eclipse occult. | 8 ± 9 | 332 ± 11 | – | – | 0.99 | – | 416 ± 21 | 295 ± 24 | 316 ± 67 | 298 ± 24 | 1148 ± 49 | 52 ± 11 | 5.1 |
| 2019-06-25 10:10 | 201906A | Acala Fluctus | Eclipse & Sunlit | 8 ± 9 | 332 ± 11 | 3.0 | 47 | 0.68 | – | 259 ± 11 | 164 ± 10 | 116 ± 41 | 199 ± 12 | 1265 ± 45 | 20 ± 3 | 2.9 |
| 2020-10-15 7:10 | 202010A | *(not constrained)* | Eclipse occult. | *(5 ± 60)* | *(20 ± 16)* | – | – | < 0.98 | 416 ± 17 | 546 ± 32 | 303 ± 32 | 183 ± 62 | 309 ± 33 | 1281 ± 37 | 37 ± 6 | 5.6 |
| 2021-08-13 11:30 | 202108A | Pillan Patera | Sunlit imaging | -9 ± 9 | *242 ± 10* | 1.5 | 26 | 0.90 | – | – | 151 ± 32 | 234 ± 164 | 182 ± 38 | – | – | – |
| 2021-08-27 10:0 | 202108D | *Pfu1410?* | Sunlit imaging | -10 ± 10 | *183 ± 13* | 1.6 | 40 | 0.76 | – | – | 301 ± 39 | 475 ± 100 | 399 ± 52 | – | – | – |

**Table 3.** Table of the seven outbursts characterized in this study. Note that the intensity, area, and power values are not corrected for emission angles. The μ column gives the constraints on the emission correction of each event, μ = 1/cos(emission).

### 4.1.1 201905A Outburst

We observed an outburst on May 8, 2019, at 14:25 Universal Time (UT), during which the total brightness of Io in eclipse increased substantially in all observed bands (Figure 2), but most dramatically in the contK-band. Using the contK-band and Lp-band disk-integrated photometry measurements, we fit a one-temperature (1T) blackbody and approximated 201905A's temperature at 1160 K with an area of 52 km² (Figure 6 and Table 3). Although we also measured Io's M-band intensity, as illustrated in Figure 6 and recorded in Table 2, we used only the contK-band and Lp-band intensities to more accurately capture the outburst's high-temperature component and to minimize the contribution of Io's many low-temperature hot spots.

On the same night, we observed Io's occultation disappearance, which provides a one-dimensional constraint for the source location (Figure 8). Direct imaging of Io in eclipse is not useful for further constraining the location, unfortunately, because the outburst fully dominates Io's emission, leaving insufficient contrast to discern Io's limb or other hot spots. Images obtained in eclipse that night, therefore, only show a single point source. In the absence of other data, we use the occultation measurements in the weeks before and after 201905A, which show a measurable hot spot, of much lower brightness than an outburst or even mini-outburst, at an occultation phase consistent with the outburst. On May 1 and 17, approximately one week before and after the May 8 outburst, we observed brightness of 6 and 8 GW/sr/µm, respectively, for this source. Months before this, on March 23 and 30, we observed similar hot spots at 15 and 10 GW/sr/µm, respectively. Although each detection is well below the threshold for outbursts or mini-outbursts, they are robustly visible in the occultation curves only after March. These hot spots are likely from the same source as the 201905A outburst and, as we will see, likely related to the next outburst observed six weeks later. Furthermore, eclipse images obtained on these nights suggest that this volcanic source is in the northern hemisphere.

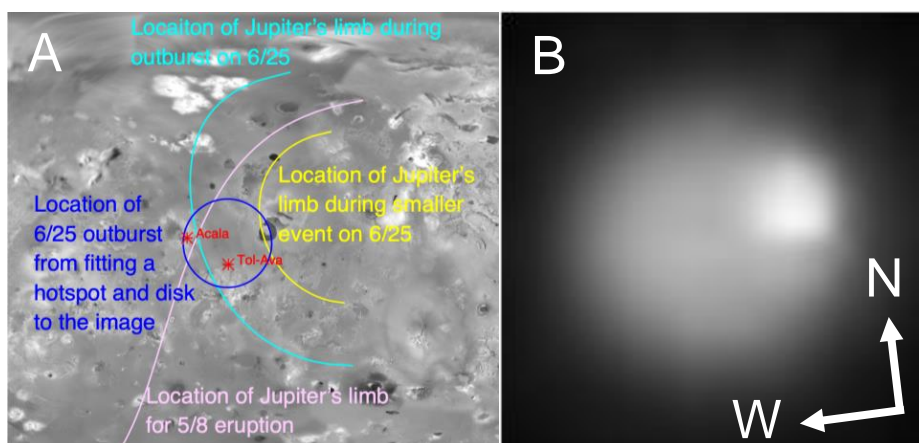

**Figure 5.** Localization constraints of 201905A and 201906A outbursts at Acala Fluces. The image in (B) is IRTF SpeX direct image of Io on June 25, 2010.

### 4.1.2 201906A Outburst

On June 25, 2019, at 10:10 UT, we observed another outburst, approximately half as bright as the 201905A event (Table 3). The 201906A exhibited intense values in the contK-band, a signature of powerful high-temperature volcanism. Again, using the contK-band and Lprime-band brightness, we estimated an effective temperature of 1214 K and a projected area of 37 km² (Figure 6 and Table 3). The 201906A event appears to be roughly half the area and similar in temperature (perhaps slightly hotter) than 201905A.

The occultation reappearance profile positions 201906A either north or southeast of 330 W (Figure 6). Because solar opposition occurred in early June, our observing geometry changed from disappearance to reappearance between eclipse observations in May and those obtained in June. The one-dimensional spatial constraint of this event notably intersects that of the May 201905A event. Because the time between Io reappearance and exit from the eclipse is shorter near solar opposition, our observation procedure allowed for additional images of Io in reflected sunlight. This is the only outburst in this campaign that we observed in all three observation modes. Because Io's limb is visible in sunlit images, we were able to constrain the outburst's location to 8±9°N and 332±11°W. Figures 5 and 8 show that this location coincides with the intersection of the two occultation constraints. Additionally, a faint hot spot persisted in every observation between May and June at occultation phases consistent with the 201906A location. It is therefore likely that 201905A and 201906A erupted from the same source region and that the outbursting hot spot was volcanically active for about 7 months between March and October, 2019.

The sunlit localized region of 201906A has several known hot spots discovered by Galileo. These include Acala Fluctus (11°N, 337°W), Tol-Ava Patera (2°N, 322°W), Fuchi Patera (28°N, 328°W) (Rathbun et al., 2004; Lopes et al., 2007), of which Acala Fluctus is the most consistent with our localization from both the sunlit imaging and the occultation curve crossing. Acala Fluctus is a dark region notable for its lava flows and a low dust content plume detected only by auroral emissions during eclipse (Geissler et al., 2004; Perry et al., 2007; Veeder et al., 2009). Tol-Ava Patera and Fuchi Patera were active both during the Galileo mission and in 2017 when they were observed from ground-based telescopes with adaptive optics (de Kleer et al., 2019). These hot spots are not associated with any previously observed outbursts or mini-outbursts. We can rule out Loki Patera (13°N, 309°W), which was near the minimum of its regular low-temperature mini-outbursting cycle (de Kleer and Rathbun, 2023). Although another possibility is that 201905A and 201906A have no relation or only weak relation to any previously identified hot spot in the region, we interpret the evidence as strongly in favor that the Acala Fluctus region produced both outbursts.

### 4.1.3 202010A Outburst

On October 15th, 2020, at 07:10 UT, we observed the most energetic high-temperature outburst of our 2016-2022 IRTF campaign. We observed 202010A in an additional methane band at 1.6 μm for a more precise measurement of the event's high-temperature emission, from which the eclipse measurements in the CH4_l-band, contK-band, and Lp-band yield an effective temperature and projected area of 1279 K and 37 km² (Figure 6 and Table 3). After a small emission angle correction, $\mu \approx 0.98$, the total power of all three events have a similar power around P ~ 5 GW. However, what distinguishes 202010A is the shape of its spectrum. The four photometric bands are consistent with a hot spot of uniform temperature. Over the range of 1.5 μm to 5 μm, 202010A does not exhibit the long-wavelength enhancement that characterizes 201905A and 201906A.

The occultation reappearance profile constraint of 202010A extends north-northwest and south-southeast of longitude 15 W (figure 8). Unlike the previous two outbursts observed in occultation, we did not see a hot spot at the same location in the weeks before and after 202010A indicating that this even appeared and disappeared quickly without a trace. Considering its single temperature spectrum, our observation of 202010A was exceptionally fortuitous. Unfortunately, we cannot constrain its location any further than the one-dimensional curve of Figure 8.

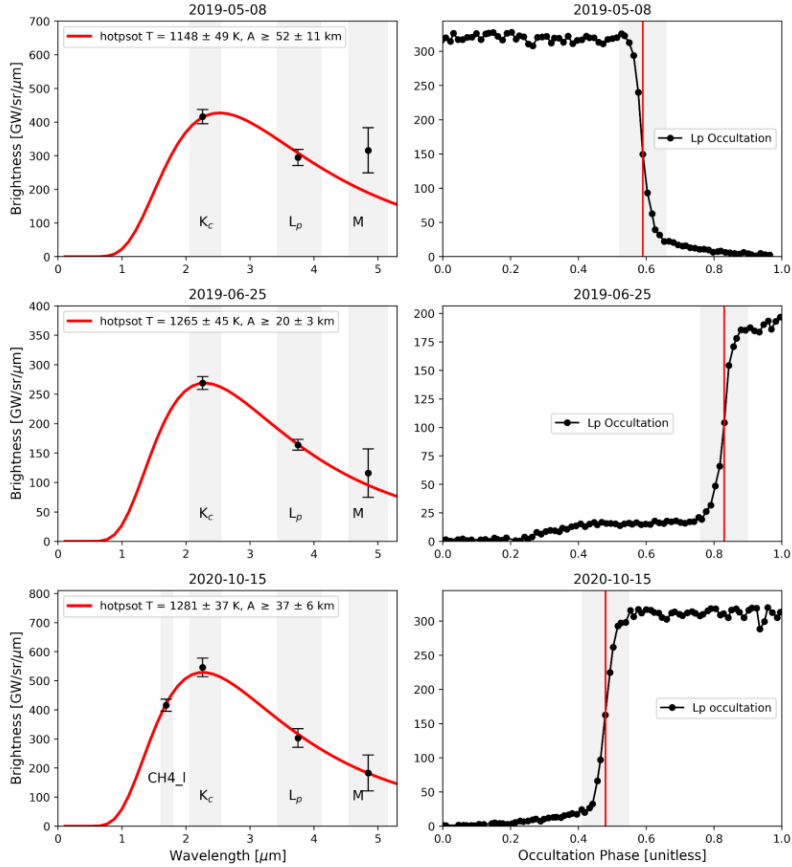

**Figure 6.** The effective temperature and area constraints for the 201905A (top), 201906A (middle), and 202010A (bottom) outburst events. The red curves on the left side are the blackbody fits of the non-corrected photometric intensity measurements: Lp-band, Kcont-band, and CH4_l-band (when available). The occultation light curve is the Lprime-band brightness of Io's observable (non-occluded) surface versus occultation phase, which is the rescaled time with 0 being the start and 1 being the finish of the occultation. The shaded regions on the left indicate the in-band wavelength range, and on the right, they indicate the transitional phase range of the outbursts.

## 4.2 Sunlit Observations

We identified four outbursts in our sunlit Lp-band observations and localized these events to approximately 15° on Io. Interestingly, three of the four outbursts were discovered at phase angles below two degrees. The uncertainty of Io's phase-curve and opposition surge adds complexity to the already challenging task of parameterizing Io's photometric properties. We primarily refrained from calculating effective temperatures for sunlit Lp-band and M-band measurements because, as evidenced by the 201905A and 201906A events, these bands are likely to contain flux from hot spots other than the outburst and inadequate for capturing either the high- or medium-temperature components of the event. Furthermore, our thermal intensity constraints in IRFT sunlit imaging have high uncertainties. Even though our capacity for scientific analysis using the sunlit photometry of Io is somewhat limited compared to eclipse observations, these sunlit data are still capable of detecting the largest volcanic events across all longitudes.

### 4.2.1 201801A Event

The first outburst we observed was on January 18, 2018, when Io was at a phase angle of alpha = 5.4°. We localized 201801A near 10±14°N 220±15°W (Table 3, Figure 7), the nearest hot spot to which is Mulungu Patera (17N 218W) (Lopes et al., 2001). Although 201801A was a powerful event with a measured intensity of $I_{Lp}$ = 183 GW/sr/µm, it did not stand out from the noise until we applied the phase correction to the sunlit time-series (Figures 4 and 7). Without this correction, this event would have likely gone undetected. From this observation, we conclude that IRTF's sensitivity to large volcanic events in sunlight Lp-band images is near our outburst threshold of $I_{Lp}$ > 150 GW/sr/µm.

### 4.2.2 UP 254W outburst, in May 2018

We observed an outburst on May 10, 2018, and localized it to a location of 36±12°S 253±15°W (Figure 7). Due to its geographical proximity, we identify this outburst as UP 254W which Gemini North observed in Lp-band and M-band approximately 4.5 hours after our observations (de Kleer et al., 2019b). Given Gemini's superior spatial, spectral, and temporal resolutions, we adopt the name, localization, and temperature-area fit of that study (de Kleer et al., 2019b). However, IRTF measured an Lp-band intensity dramatically higher than Gemini North, which we interpret as a physical change in the volcano's activity during the 4.5 hours between the observations. Although the de Kleer et al., (2019a) identified this source as a mini-outburst, the earlier IRTF measurement places the event well within the Lp brightness range of other outbursts.

Gemini North observed Io at 15:15 UT with an Lp-band intensity at UP 254W $I_{Lp}$= 134 ± 24 GW/sr/µm, and IRTF observed Io at 10:45 UT with $I_{Lp}$= 275 ± 99 GW/sr/µm. In 4.5 hours between these observations, the intensity of UP 254W seemed to drop by a factor of two. Despite the high error of the IRTF measurement, these intensities are separated by a little more than 1-sigma. This discrepancy can be explained in two ways: as either a calibration error or a rapid decay in the thermal intensity of the volcano. Despite the uncertainties in our photometric phase-curve correction and other differences between the IRTF and Gemini calibration pipelines, this magnitude of change is too large to explain solely from the differences between these two observation methods. Given that Gemini North and IRTF are both on Mauna Kea, which had clear weather throughout the May 10, 2018, observation night, we can eliminate atmospheric transparency changes as the cause. Similarly, both observations had the same low phase angle. If there were a major systematic difference in these methods, it would likely be from the annulus background subtraction used to estimate the hot spot intensities in (de Kleer and de Pater, 2016a; de Kleer et al., 2019b), which could underestimate the hot spot's intensity if the rest of Io's surface is abnormally bright due to its opposition surge. The phase angle on May 10, 2018, was 0.4°, which is within the surge angle of several photometric models of Io's surface (Simonelli and Veverka 1986; Shavlovskij, 2005). The values for previous sunlit phase angles span from 0.5 to 12°, making this one of the lowest phase angles in our campaign.

In the absence of evidence to the contrary, we assume that the differing Lp-band measurements are due to a real change in the thermal intensity of the UP 254W volcanic event. This outburst provides a valuable opportunity to observe the short-timescale evolution of an outburst. This change in Lp-band intensity from 275 to 134 GW/sr/µm in only several hours could be the result of either rapidly cooling effusive lava or subsidence in the flow rate of fire fountains, both of which were observed at previous outbursts (Davies et al., 2001, Davies et al., 2005).

### 4.2.3 August 2021 Events

Io exhibited high levels of volcanic activity in August 2021, characterized by two outbursts: 202108A and 202108D, which occurred on August 13 and 27, 2021, respectively. The Lp-band and M-band images in Figure 7 spotlight these localized hot spots. Unlike in previous IRTF observations in sunlit mode, we were able to capture both the M-band intensity and the customary Lp-band, providing not only temperature estimates but, more importantly, a second band for hot spot localization on Io. The M-band images enhance hot spot localization by offering a more favorable ratio between reflected and emitted light.

The IRTF captured images of Io in reflected sunlight seven times between August 13 and September 2, encompassing the entire equatorial region (Figure 3). Out of these seven observation nights, four—August 14, August 19, August 21, and September 2—did not exhibit unusually high thermal activity. These nights of non-detection surveyed the leading and sub-Jovian hemispheres, extending from approximately 150°W to 330°W longitude. The other three observation nights—August 13, 26, and 27—detected hot spots exclusively on the trailing hemisphere, between 20°N and 30°S (Figure 7).

On August 13, two distinguishable hot spots, 202108A and 202108B, were observed (third row of Figure 7). The former, brighter hot spot, was an outburst near Pillan Patera (9±9°S 242±10°W), and the latter, a dimmer hot spot, was likely Loki Patera, which was known to be active several weeks earlier(de Kleer and Rathbun, 2023). Despite significant observational uncertainties, the outburst near Pillan aligns with Pillan's recorded location variability (de Pater et al., 2015). The August 27 images faintly showed a secondary hot spot, 202108E, near 10°S 240°W, which could be the afterglow of the 202108A Pillan outburst from two weeks earlier (bottom row of Figure 7). While there may be a very faint glow in Lp-band, there is clearly activity in the M-band. While it is not bright enough for us to calculate the total brightness, the temperature of this area has clearly decreased.

On August 27, Io exhibited the most substantial outburst of our IRTF campaign, 202108D, with an intensity of $I_{Lp}$ = 402 GW/sr/μm near 10±10°S and 183±13°W. Although this region of Io is remote from any previously observed outburst or mini-outburst (Veeder et al., 2012, Cantrall et al., 2018), it correlates with an undivided patera floor Pfu1410 (6.1°S 186.4°W) (Williams et al., 2011), which was not observed as a hot spot until 2010 when Keck observed it emit a modest 0.5 GW/sr/micron in the Lprime-band (de Pater et al., 2014a). The closest hot spots identified by Galileo to 202108D are Aidne Patera (2°S, 178°W) and Haokah (19S 185W) (Lopes et al., 2004). Neither hot spot seems a likely candidate for an outburst, and it is plausible that the 202108D outburst originated from a new or dormant location. By the time we observed this location again on September 22 and 24, Io's total brightness had returned to nominal values with no apparent hot spot.

Intriguingly, each of the five hot spots identified in August 2021 was located within the trailing hemisphere and near the equator. Both the 202108A and 202108D outbursts were likely transient events, without intensities large enough to be detectable in the sunlit images before or after. In the fourteen days between August 13 and 27, we had four other observation nights with sub-observer longitudes in the leading hemisphere and were unfortunately unable to observe the outburst locations. However, given that both sunlit observations of the trailing hemisphere revealed outbursts, it is plausible that this area of Io was in a state of heightened activity during this period.

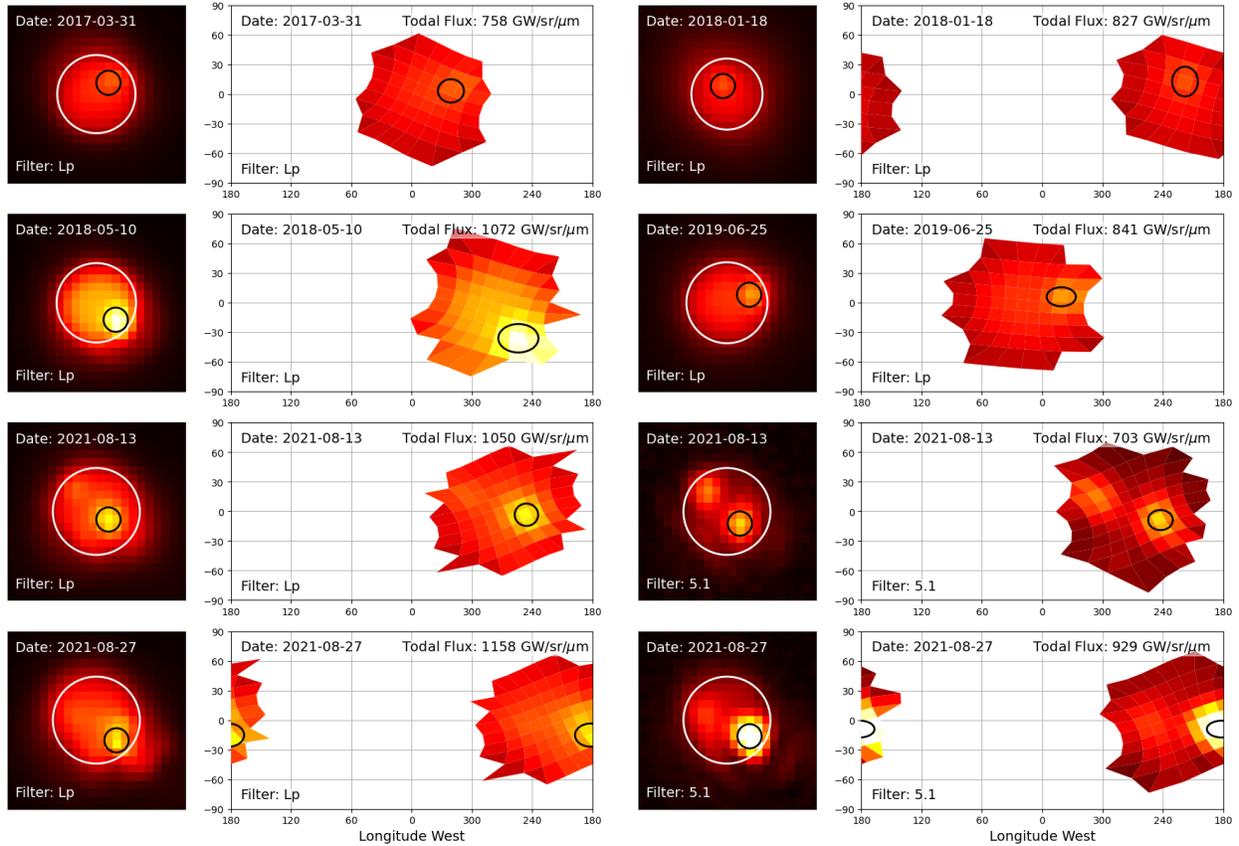

**Figure 7.** Sunlit IRTF images of Io on several nights that show clear hot spots. The square "shift and add" images are used to constrain spot location and project each pixel onto the map of Io that is to the right of each image. The heatmap scale is the same on all images. The bottom two rows show Lp (left) and M (right) images from August when both bands are imaged. All circled hot spots are outbursts, except for the upper left panel for 2017-03-31, in which Loki is a mini-outburst.

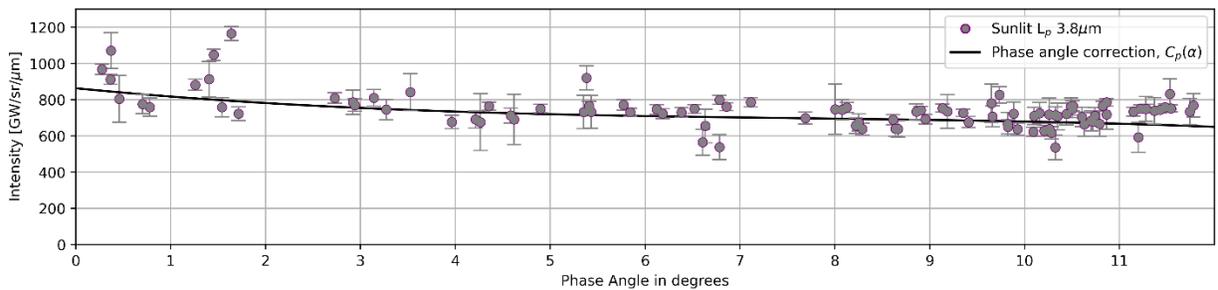

**Figure 8.** The phase angles of Io's total brightness values for the sunlit observations. Of the full range from 0° to 12°, there are twelve sunlit observations with phase angles below 2°. These low phase angle

observations in reflected sunlight may have a significant brightness contribution from the opposition surge effect.

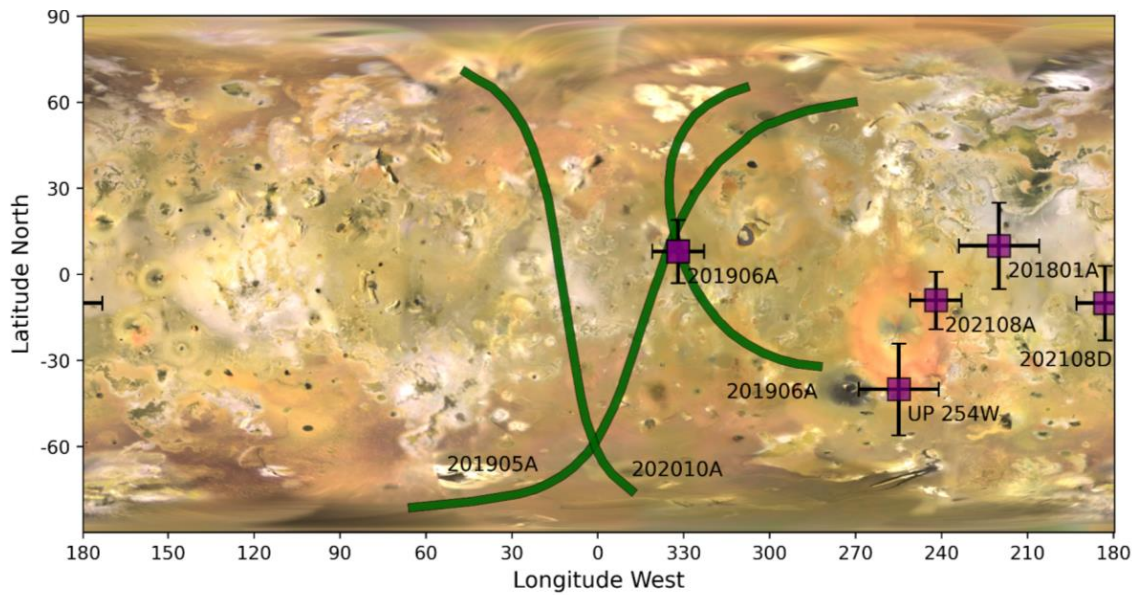

**Figure 9.** Map showing the constrained locations of the outbursts characterized in this paper. The green lines are the occultation detections, and the outburst can be anywhere along the line. The purple squares show the outbursts discovered in sunlight; the black lines are their localization uncertainties.

# 5 Discussion

The frequency of outbursts on Io is a crucial constraint on their role in Io's heat cycle and flux balance. During our IRTF campaign, we observed seven events over about 170 observation nights, yielding a probability of observing an outburst at 4.0 ± 1.5%. Remarkably, this value aligns with the pre-Galileo estimate of 3.3 ± 1.5% (Spencer and Schneider 1996), suggesting that the occurrence of outbursts has been stable over decadal timescales. However, our detections, along with previous studies, show that outbursts can cluster on monthly or yearly timescales. This was clearly illustrated by back-to-back outbursts in August 2021 and May-June 2019. Previous studies also observed this behavior, such as the Heno Patera, Rarog Patera, and 201308C outbursts that occurred within two weeks, followed by several years with no detected outbursts (de Kleer and de Pater 2016a, de Kleer et al., 2019b).

During our IRTF study period, 2016-2022, outbursts appeared to favor the trailing hemisphere. However, the spatial distribution of the five sunlit outbursts was different from the three found in eclipse. With the exception of the events at Acala Fluctus, the sunlit outbursts occurred in the vicinity of the anti-jovian point between 260W to 170W. The eclipse outbursts were constrained between 70W and 270W. The eclipse observation mode is, by necessity, centered on the sub-Jovian point within several degrees and cannot detect activity in the region of the four outbursts between 70W and 270W discovered in reflected sunlight. Although the sunlit mode is theoretically unbiased in longitude, our IRTF coverage is not uniform, as seen in the histogram in the right subplot of Figure 3. The total number of observations with sub-observer longitudes between the quadrant 270W and 180W is 15, which is in fact lower than the average of 25 expected from uniform coverage even though more outbursts are observed in this area. If we adjust the estimated frequency of outbursts to account for both the preference for 270-180W longitudes and the observational bias, then this region of Io could have been outbursting 27 ± 13% of the time. This level of activity from 2018 to 2021 is substantially greater than the total outburst frequency of 4 ± 1.5%.

The volcanic activity on Io in August 2021 demonstrated an eastward migration, reminiscent of the activity observed between 2013 and 2015 (de Kleer and de Pater, 2016b). During that period, a consistent eastward progression of bright eruptions (including outbursts and mini-outbursts) occurred in the southern hemisphere, between 300 and 200W, over two years. Our observations reveal a similar phenomenon happening near the equator within about a month. The activity in August 2021 may have started with Loki Patera, which, in our sunlit images, peaked sometime between May and August 2021, and extended to Pillan Patera and further east to a new outburst location near Aidne Patera at the end of August. We observed a similar phenomenon happening near the equator within the timeframe of a month. De Kleer and de Pater(2016b) proposed a large-scale volcanic triggering mechanism to explain this, but testing this hypothesis proves challenging.

A significant finding from our observations is that outbursts can recur at a given location. Prior to this study, Tvashtar was the only location unequivocally identified as experiencing repeated outbursts, with approximately six instances between 1999 and 2007 (Veeder et al., 2012; Cantrall et al., 2018). We present persuasive evidence of two additional locations with recurring outbursts. The Acala Fluctus region manifested outbursts in May and June of 2019, and an outburst at Pillan Patera occurred in August 2021, following its last recorded outburst in 1997 (Davies et al., 2007; Keszthelyi et al., 2007; Howell et al., 2001; Williams et al., 2001). Although Pillan Patera was observed to have as many as four mini-outbursts between 2007 and 2015 (de Pater 2016a; Lellouch et al., 2015; de Pater et al., 2016), it was not clear that any of these qualified as an outburst. With our recent observations, we can add two more locations to the category of recurring outbursts, increasing this count from one to three.

While some outbursts recur, we have also identified at least two other outbursts from novel, unexpected locations. It's noteworthy that we cannot associate the locations of the 201801A and 202108D outbursts with Io's known hot spots, as identified by the Galileo mission and subsequent surveys. Consequently, our understanding of the where, when, and why Io's outbursts occur remains enigmatic.

# 6 Conclusions

We present five years of near-infrared measurements of Io, acquired with IRTF from 2016 to 2022. Our analysis reveals seven volcanic outbursts, substantially increasing the number of known outbursts for more comprehensive studies of this rare eruption style. The only mini-outbursts identified in this study likely originated from Loki Patera.

We also present a new and simplified definition of an outburst based on its thermal emission in the Lp-band (3.8 μm) instead of using the somewhat outdated definition that requires an outburst to double Io's 5-μm brightness. Our definition requires reasonable evidence of two criteria. A thermal detected outburst must have an intensity above the threshold $I_{3.8\mu m} > 150$ GW/sr/μm and be localized to a confined region on Io.

Three of the seven outbursts were discovered in eclipse and occultation photometry of Io's sub-Jovian hemisphere. The absence of reflected sunlight enables more precise constraints of the temperature and area of each event. Spatial constraints for these outbursts are limited to the one-dimensional projection of Jupiter's limb on Io (Figure 8). The sharp intensity changes caused by these outbursts (Figure 6) sufficiently demonstrate that the outburst is constrained on Io's surface, although we cannot further constrain its location.

The absence of detections at this approximate location before and after the 202010A event illustrates that it was an exceptionally brief, transient event. Since it had not been erupting long, we would not expect there to be substantial amounts of cooling lava flows. The fact that our single-temperature fit (Figure 6) fits all four observed wavelengths to within the error bars is consistent with this eruption being young when we observed the outburst. This contrasts with 201905A and 201906A outbursts at Acala Fluctus, which was likely a more developed eruption at the time of detection with large areas of cooling lava flows that remained detectable for months.

The remaining four outbursts were observed only in reflected sunlight. Unlike eclipse observations, sunlit observations can detect bright hot spots at any longitude. Notably, all four outbursts detected between 2018 and 2021 were in the trailing hemisphere, even though we observed all longitudes. This cannot be due to an observational bias, as the right panel of Figure 3 illustrates that we imaged the opposite or leading hemisphere more than the one where we detected the outbursts.

The UP 254W event in May 2018 was a rare instance of observing an outburst fading into a mini-outburst. Although its thermal characterization is limited to two photometric bands, the rapid decrease in intensity by a factor of two over 4.5 hours is most likely due to changes in its high-temperature fire fountains.

Io exhibited high activity in August 2021, with two outbursts and several potential mini-outbursts clustered near the equator of the trailing hemisphere. Leading hemisphere observations during this period did not reveal any hot spots. The highly clustered nature of these eruptions is reminiscent of the 2013 Rarog Patera and Heno Patera outbursts (de Pater 2014b, de Kleer and de Pater 2016b).

Our observations imply several significant conclusions. Firstly, we found that outbursts can recur at a given location, a characteristic previously only attributed to the Tvashtar location. Both the Acala Fluctus region and Pillan Patera experienced recurrent outbursts, expanding this class of outburst from one to three. Secondly, we have identified outbursts emanating from new and unexpected locations.

Overall, this campaign demonstrates that medium-class telescopes like IRTF can significantly contribute to the study of Io's volcanism. Considering the importance of regularly monitoring Io's ever-changing activity, it is crucial for ground-based observation campaigns to continuously observe Io year after year. The value of these data will likely increase with the baseline over which they allow us to analyze Io's complex behavior.

# 7 References

Blaney, D., Johnson, T., Matson, D. L., Veeder, G. J. (1995). Volcanic Eruptions on Io: Heat Flow, Resurfacing, and Lava Composition. *Icarus*, *113*(1), 220–225. https://doi.org/10.1006/icar.1995.1020

Blaney, D. L., Veeder, G. J., Matson, D. L., Johnson, T. V., Goguen, J. D., & Spencer, J. R. (1997). Io's thermal anomalies: Clues to their origins from comparison of ground based observations between 1 and 20 µm. *Geophysical Research Letters*, *24*(20), 2459–2462. https://doi.org/10.1029/97GL02509

Carr, M. H. (1986). Silicate volcanism on Io. *Journal of Geophysical Research: Solid Earth*, *91*(B3), 3521–3532. https://doi.org/10.1029/jb091ib03p03521

Cantrall, C., de Kleer, K., de Pater, I., Williams, D. A., Davies, A. G., and Nelson, D. (2018). Variability and geologic associations of volcanic activity in 2001–2016. *Icarus*, *312*, 267–294. https://doi.org/10.1016/j.icarus.2018.04.007

Davies, A. G., Keszthelyi, L. P., and Harris, A. J. L. (2010). The thermal signature of volcanic eruptions on Io and Earth. *Journal of Volcanology and Geothermal Research*, *194*(4), 75–99. https://doi.org/10.1016/j.jvolgeores.2010.04.009

de Pater, I., Davies, A. G., Ádámkovics, M., & Ciardi, D. R. (2014). Two new, rare, high-effusion outburst eruptions at Rarog and Heno Paterae on Io. *Icarus*, *242*, 365–378. https://doi.org/10.1016/j.icarus.2014.06.016

de Pater, I., Laver, C., Davies, A. G., de Kleer, K., Williams, D. A., Howell, R. R., Rathbun, J. A., & Spencer, J. R. (2016). Io: Eruptions at Pillan, and the time evolution of Pele and Pillan from 1996 to 2015. *Icarus*, *264*, 198–212. https://doi.org/10.1016/j.icarus.2015.09.006

de Pater, I., de Kleer, K., Davies, A. G., & Ádámkovics, M. (2017). Three decades of Loki Patera observations. *Icarus*, *297*, 265–281. https://doi.org/10.1016/j.icarus.2017.03.016

de Kleer, K., and de Pater, I. (2016a). Time variability of Io's volcanic activity from near-IR adaptive optics observations on 100 nights in 2013–2015. *Icarus*, *280*, 378–404. https://doi.org/10.1016/j.icarus.2016.06.019

de Kleer, K., and de Pater, I. (2016b). Spatial distribution of Io's volcanic activity from near-IR adaptive optics observations on 100 nights in 2013–2015. *Icarus*, *280*, 405–414.
https://doi.org/10.1016/j.icarus.2016.06.018

de Pater, I., de Kleer, K., Davies, A. G., & Ádámkovics, M. (2017). Three decades of Loki Patera observations. Icarus, 297, 265–281. https://doi.org/10.1016/j.icarus.2017.03.016

de Kleer, K., Nimmo, F., and Kite, E. (2019a). Variability in Io's Volcanism on Timescales of Periodic Orbital Changes. *Geophysical Research Letters*, *46*(12), 6327–6332. https://doi.org/10.1029/2019GL082691

de Kleer, K., de Pater, I., Molter, E. M., Banks, E., Davies, A. G., Alvarez, C., Campbell, R., Aycock, J., Pelletier, J., Stickel, T., Kacprzak, G. G., Nielsen, N. M., Stern, D., and Tollefson, J. (2019b). Io's Volcanic Activity from Time Domain Adaptive Optics Observations: 2013–2018. *The Astronomical Journal*, *158*(1), 29. https://doi.org/10.3847/1538-3881/ab2380

de Kleer, K., & Rathbun, J. A. (2023). Io's Thermal Emission and Heat Flow. *Io: A New View of Jupiter's Moon*, 173-209.

Howell, R. R., Spencer, J. R., Goguen, J. D., Marchis, F., Prangé, R., Fusco, T., ... & Hege, E. K. (2001). Ground-based observations of volcanism on Io in 1999 and early 2000. Journal of Geophysical Research: Planets, 106(E12), 33129-33139.

Lellouch, E., Ali-Dib, M., Jessup, K. L., Smette, A., Käufl, H. U., & Marchis, F. (2015). Detection and characterization of Io's atmosphere from high-resolution 4-μm spectroscopy. *Icarus*, *253*, 99-114.

Lopes, R. M. C., Kamp, L. W., Douté, S., Smythe, W. D., Carlson, R. W., McEwen, A. S., Geissler, P. E., Kieffer, S. W., Leader, F. E., Davies, A. G., Barbinis, E., Mehlman, R., Segura, M., Shirley, J., & Soderblom, L. A. (2001). Io in the near infrared: Near-Infrared Mapping Spectrometer (NIMS) results from the Galileo flybys in 1999 and 2000. *Journal of Geophysical Research: Planets*, *106*(E12), 33053–33078. https://doi.org/10.1029/2000JE001463

Lopes, Rosaly MC, and David A. Williams. "Io after Galileo." Reports on Progress in Physics 68.2 (2005): 303.

Lopes, R. M. C., de Kleer, K., Tuttle, J., & Editors, K. (n.d.). *Io: A New View of Jupiter's Moon Second Edition*. (2023)

Keszthelyi, L., McEwen, A. S., Phillips, C. B., Milazzo, M., Geissler, P., Turtle, E. P., Radebaugh, J., Williams, D. A., Simonelli, D. P., Breneman, H. H., Klaasen, K. P., Levanas, G., Denk, T., Alexander, D. D. A., Capraro, K., Chang, S. H., Chen, A. C., Clark, J., Conner, D. L., … Pilcher, C. B. (2001). Imaging of volcanic activity on Jupiter's moon Io by Galileo during the Galileo Europa Mission and the Galileo Millennium Mission. *Journal of Geophysical Research: Planets*, *106*(E12), 33025–33052. https://doi.org/10.1029/2000JE001383

Keszthelyi, L., Jaeger, W., Milazzo, M., Radebaugh, J., Davies, A. G., & Mitchell, K. L. (2007). New estimates for Io eruption temperatures: Implications for the interior. *Icarus*, *192*(2), 491–502. https://doi.org/10.1016/j.icarus.2007.07.008

Keszthelyi, L. P., & Suer, T. A. (2023). The Composition of Io. In Io: A New View of Jupiter's Moon (pp. 211-232). Cham: Springer International Publishing.

Marchis, F., de Pater, I., Davies, A. G., Roe, H. G., Fusco, T., le Mignant, D., Descamps, P., Macintosh, B. A., and Prangé, R. (2002). High-resolution Keck adaptive optics imaging of violent volcanic activity on Io. *Icarus*, *160*(1), 124–131. https://doi.org/10.1006/icar.2002.6955

McEwen, A.S., Keszthelyi, L., Spencer, J.R., Schubert, G., Matson, D.L., Lopes-Gautier, R., Klaasen, K.P., Johnson, T.V., Head, J.W., Geissler, P. and Fagents, S. (1998). High-temperature silicate volcanism on Jupiter's moon Io. Science, 281(5373), 87-90.

Milazzo, M. P., Keszthelyi, L. P., Radebaugh, J., Davies, A. G., Turtle, E. P., Geissler, P., Klaasen, K. P., Rathbun, J. A., and McEwen, A. S. (2005). Volcanic activity at Tvashtar Catena, Io. *Icarus*, *179*(1), 235–251. https://doi.org/10.1016/j.icarus.2005.05.013

Rathbun, J. A., and Spencer, J. R. (2010). Ground-based observations of time variability in multiple active volcanoes on Io. *Icarus*, *209*(2), 625–630. https://doi.org/10.1016/j.icarus.2010.05.019

Shavlovskij, V. I. (2005). Opposition effects of Jupiter's satellites Io and Europa. *Кинематика и физика небесных тел*.

Simonelli, D. P., & Veverka, J. (1986). Phase curves of materials of Io: Interpretation in terms of Hapke's function. *Icarus*, *68*(3), 503-521.

Sinton, W. M., Lindwall, D., Cheigh, F., and Tittemore, W. C. (1983). Io: The Near-Infrared Monitoring Program, 1979-1981. In *ICARUS* (Vol. 54).

Spencer, J. R., and Schneider, N. M. (1996). IO ON THE EVE OF THE GALILEO MISSION. In *Annu. Rev. Earth Planet. Sci* (Vol. 24). www.annualreviews.org

Tokunaga, A. T., and Vacca, W. D. (2005). The Mauna Kea Observatories Near-Infrared Filter Set. III. Isophotal Wavelengths and Absolute Calibration. *Publications of the Astronomical Society of the Pacific*, *117*(830), 421–426. https://doi.org/10.1086/429382

Veeder, G. J., Davies, A. G., Matson, D. L., Johnson, T. v., Williams, D. A., and Radebaugh, J. (2012). Io: Volcanic thermal sources and global heat flow. *Icarus*, *219*(2), 701–722. https://doi.org/10.1016/j.icarus.2012.04.004\

# 8 Appendix

**Table 1s.** List of the eclipse observations used in this analysis. The sub-observer longitude and latitude are given, as are the measured intensity and uncertainty values in units of GW/sr/ μm for each photometric band used.

| date and time | longitude West | latitude North | I CH4_l | I CH4_l sigma | I contK | I contK sigma | I Lp | I Lp sigma | I M | I M sigma |
|---|---|---|---|---|---|---|---|---|---|---|

| Date | | | | | | | | | | |
|---|---|---|---|---|---|---|---|---|---|---|
| 2016-11-27 15:14 | 347.7 | -2.6 | | | | | 19.3 | 6.9 | | |
| 2017-01-12 15:28 | 346.5 | -2.9 | | | 5.0 | 3.4 | 19.3 | 7.8 | | |
| 2017-02-04 15:31 | 344.5 | -3.0 | | | 5.2 | 0.8 | 21.8 | 5.6 | | |
| 2017-02-13 11:59 | 345.0 | -3.0 | | | 5.5 | 1.9 | 23.9 | 4.8 | | |
| 2017-02-27 15:48 | 348.5 | -3.1 | | | | | 35.0 | 6.7 | | |
| 2017-05-09 12:05 | 12.2 | -2.9 | | | | | 62.1 | 5.3 | | |
| 2017-05-11 06:44 | 10.5 | -2.9 | | | 13.1 | 2.3 | 63.7 | 4.3 | 153.3 | 47.6 |
| 2017-05-25 10:15 | 6.3 | -2.8 | | | 14.2 | 2.3 | 75.0 | 3.8 | 168.0 | 13.6 |
| 2017-06-26 06:44 | 13.9 | -2.6 | | | 8.4 | 1.6 | 57.1 | 7.5 | 201.7 | 120.0 |
| 2017-07-19 06:44 | 12.1 | -2.6 | | | | | 37.9 | 5.8 | | |
| 2017-08-13 02:03 | 17.5 | -2.6 | | | | | 33.8 | 8.5 | | |
| 2018-01-06 21:53 | 345.8 | -3.2 | | | 3.9 | 1.5 | 36.4 | 8.1 | 111.1 | 53.0 |
| 2018-01-08 16:24 | 345.3 | -3.2 | | | 5.1 | 0.8 | 22.1 | 3.0 | 34.1 | 10.0 |
| 2018-01-15 18:23 | 346.9 | -3.2 | | | 2.5 | 0.9 | 22.9 | 2.3 | 51.5 | 10.3 |
| 2018-04-12 11:26 | 341.8 | -3.4 | | | | | 34.0 | 6.1 | | |
| 2018-04-26 14:50 | 348.1 | -3.4 | | | 4.9 | 2.3 | 15.1 | 5.9 | | |
| 2018-05-23 06:00 | 7.8 | -3.3 | | | 13.5 | 3.0 | 93.5 | 8.6 | | |
| 2018-06-06 09:33 | 3.9 | -3.3 | | | 8.0 | 0.4 | 35.6 | 2.1 | 77.1 | 22.2 |
| 2018-06-15 05:44 | 10.4 | -3.2 | 14.6 | 7.8 | 7.7 | 0.3 | 27.0 | 1.2 | 180.2 | 26.0 |
| 2018-06-22 07:35 | 4.8 | -3.2 | | | 10.9 | 0.9 | 44.6 | 3.0 | | |
| 2018-06-29 09:27 | 7.1 | -3.2 | | | 10.0 | 0.9 | 39.6 | 2.1 | 67.8 | 12.7 |
| 2018-07-31 05:33 | 7.7 | -3.0 | | | | | 59.4 | 3.4 | | |
| 2019-02-28 14:10 | 348.7 | -2.9 | | | 10.3 | 1.1 | 39.1 | 2.2 | | |
| 2019-03-23 14:33 | 354.1 | -2.8 | | | 11.9 | 0.4 | 74.3 | 2.1 | | |
| 2019-03-30 15:58 | 347.4 | -2.8 | | | 11.7 | 1.1 | 58.8 | 14.6 | | |
| 2019-05-01 12:21 | 347.6 | -2.8 | | | 5.8 | 0.3 | 41.8 | 5.9 | | |
| **2019-05-08 14:25** | 347.7 | -2.8 | | | 425.0 | 21.0 | 327.5 | 24.2 | 418.8 | 66.8 |
| 2019-05-17 10:54 | 348.2 | -2.8 | | | | | 28.1 | 1.5 | | |
| 2019-05-24 12:27 | 344.9 | -2.8 | | | 5.6 | 0.4 | 35.2 | 2.2 | 23.2 | 9.6 |
| **2019-06-25 10:59** | 14.7 | -2.8 | | | 267.3 | 11.3 | 196.3 | 8.6 | 218.7 | 41.3 |
| 2019-07-04 07:19 | 15.1 | -2.8 | | | 7.3 | 0.5 | 28.0 | 2.0 | | |
| 2019-07-11 09:05 | 14.3 | -2.8 | | | 7.3 | 0.4 | 27.1 | 2.0 | | |
| 2019-08-28 03:57 | 15.1 | -2.6 | | | 6.8 | 0.4 | 25.4 | 1.6 | | |
| 2019-09-11 07:26 | 14.3 | -2.6 | | | 6.8 | 0.5 | 23.4 | 3.3 | | |
| 2019-09-27 05:43 | 13.8 | -2.5 | | | 8.0 | 0.5 | 29.7 | 4.0 | 101.9 | 25.3 |
| 2019-10-13 04:42 | 16.9 | -2.5 | | | 7.1 | 0.6 | 22.9 | 5.3 | | |
| 2019-10-29 03:06 | 17.5 | -2.4 | | | | | 22.1 | 5.4 | | |
| 2020-02-15 19:05 | 347.5 | -1.8 | | | 21.6 | 10.2 | 94.2 | 6.2 | | |
| 2020-02-22 20:51 | 345.4 | -1.8 | | | 17.5 | 1.1 | 113.0 | 5.1 | 238.3 | 58.8 |
| 2020-05-12 12:12 | 345.2 | -1.3 | | | 12.0 | 2.1 | 48.2 | 14.0 | 114.3 | 61.9 |
| 2020-05-19 14:17 | 348.7 | -1.3 | | | 8.2 | 0.3 | 52.1 | 6.9 | 142.1 | 40.4 |
| 2020-05-26 15:42 | 341.4 | -1.3 | | | 13.1 | 2.6 | 46.2 | 15.2 | | |
| 2020-06-11 13:54 | 347.2 | -1.3 | 8.5 | 3.9 | 6.1 | 2.9 | 21.1 | 8.4 | 151.4 | 46.3 |

| date and time | | | | | | | | | |
|---|---|---|---|---|---|---|---|---|---|
| 2020-06-20 10:15 | 343.1 | -1.3 | 5.9 | 2.0 | 7.2 | 1.0 | 33.8 | 5.2 | | |
| 2020-06-27 12:06 | 341.8 | -1.3 | 10.9 | 3.1 | 6.4 | 0.6 | 36.5 | 4.5 | 119.4 | 19.8 |
| 2020-08-07 06:56 | 12.7 | -1.3 | 9.6 | 1.3 | 20.3 | 3.4 | 20.4 | 4.6 | | |
| 2020-08-14 08:49 | 13.9 | -1.3 | 6.1 | 2.3 | 7.7 | 1.8 | 25.8 | 3.9 | 87.8 | 24.2 |
| 2020-08-21 10:35 | 13.3 | -1.3 | 8.5 | 0.7 | 8.0 | 0.9 | 34.2 | 4.4 | | |
| 2020-08-23 05:19 | 12.1 | -1.3 | 4.0 | 1.0 | 7.8 | 0.9 | 23.6 | 3.4 | 80.4 | 30.7 |
| 2020-08-30 06:52 | 11.5 | -1.4 | 7.0 | 1.1 | 9.0 | 1.0 | 23.6 | 2.8 | 83.8 | 17.1 |
| 2020-09-06 08:46 | 13.0 | -1.3 | 8.2 | 1.1 | 8.1 | 2.5 | 24.2 | 3.3 | 74.4 | 17.0 |
| 2020-09-22 07:00 | 12.7 | -1.3 | 4.9 | 2.2 | 8.4 | 1.4 | 27.4 | 4.4 | 90.8 | 26.3 |
| 2020-10-08 05:27 | 15.0 | -1.3 | 4.1 | 1.7 | 8.8 | 1.9 | 28.9 | 1.4 | 76.9 | 18.1 |
| **2020-10-15 07:05** | 12.3 | -1.3 | 415.9 | 16.5 | 551.4 | 32.4 | 325.3 | 31.5 | 257.3 | 70.0 |
| 2020-10-31 05:39 | 14.3 | -1.2 | 8.5 | 3.4 | 7.3 | 2.3 | 33.5 | 2.9 | 102.4 | 31.6 |
| 2020-12-09 04:34 | 15.5 | -1.0 | | | 7.5 | 1.3 | 22.6 | 3.6 | 64.7 | 18.2 |
| 2021-04-06 17:00 | 343.4 | 0.2 | | | 21.8 | 0.5 | 66.8 | 2.2 | 166.9 | 17.6 |
| 2021-04-13 18:43 | 347.8 | 0.2 | | | | | 63.3 | 11.0 | | |
| 2021-04-22 15:26 | 347.1 | 0.3 | | | 12.0 | 2.2 | 58.9 | 5.4 | 142.7 | 30.9 |
| 2021-05-15 15:49 | 344.5 | 0.5 | | | 14.5 | 2.0 | 78.4 | 8.2 | 205.7 | 37.3 |
| 2021-05-31 13:42 | 348.7 | 0.7 | | | 11.8 | 0.5 | 69.5 | 3.7 | 189.2 | 32.2 |
| 2021-06-07 15:32 | 345.6 | 0.7 | | | 13.3 | 0.5 | 77.1 | 5.1 | 203.3 | 21.9 |
| 2021-06-16 12:14 | 345.4 | 0.8 | | | | | 55.5 | 6.5 | 117.1 | 36.9 |
| 2021-06-23 13:46 | 345.9 | 0.8 | | | 22.0 | 1.2 | 113.0 | 5.5 | 235.0 | 47.4 |
| 2021-09-02 10:36 | 16.6 | 0.8 | | | 8.8 | 7.5 | 63.8 | 2.6 | 105.1 | 34.6 |
| 2021-09-09 12:28 | 13.0 | 0.7 | | | 7.5 | 0.9 | 37.0 | 2.6 | 97.5 | 17.1 |
| 2021-09-18 08:50 | 15.2 | 0.7 | | | | | 35.2 | 3.2 | 93.9 | 22.8 |
| 2021-10-11 08:40 | 13.4 | 0.6 | | | 8.8 | 1.5 | 33.1 | 3.2 | 98.1 | 20.8 |
| 2021-10-20 04:58 | 13.4 | 0.6 | | | 6.4 | 1.7 | 30.3 | 3.2 | 31.9 | 11.3 |
| 2021-10-27 06:54 | 13.3 | 0.6 | | | 7.9 | 0.7 | 30.1 | 1.5 | 82.3 | 20.0 |
| 2022-01-13 04:34 | 14.7 | 0.9 | | | | | 30.6 | 2.8 | 80.1 | 23.5 |
| 2022-05-11 16:44 | 343.3 | 2.0 | | | | | 29.0 | 4.6 | | |
| 2022-05-27 15:39 | 347.7 | 2.2 | | | 12.9 | 2.1 | 30.3 | 2.1 | 99.2 | 20.0 |
| 2022-06-19 15:30 | 341.4 | 2.4 | | | 12.7 | 0.9 | 32.6 | 5.3 | 112.3 | 10.3 |
| 2022-07-12 15:42 | 347.0 | 2.5 | | | 22.7 | 2.4 | 38.2 | 1.3 | 103.7 | 21.8 |
| 2022-07-21 12:09 | 344.2 | 2.6 | | | 20.3 | 1.1 | 33.3 | 5.8 | 98.8 | 29.3 |
| 2022-07-28 13:51 | 344.0 | 2.6 | | | 15.9 | 1.6 | 33.2 | 1.6 | 74.0 | 19.1 |

Table 2s. List of the sunlit observations used in this analysis. The phase angle, sub-observer longitude and latitude and measured intensity and uncertainty values in units of GW/sr/µm for each photometric band used.

| date and time | phase angle | longitude West | latitude North | I Lp | I Lp sigma | I_M | I M sigma |
|---|---|---|---|---|---|---|---|
| 2016-06-19 05:24 | 10.5 | 293 | -1.6 | 763 | 47 | | |
| 2016-12-22 15:29 | 9.8 | 32 | -2.8 | 650 | 39 | | |
| 2016-12-24 15:16 | 9.9 | 77 | -2.8 | 637 | 34 | | |
| 2017-01-02 15:37 | 10.3 | 111 | -2.8 | 638 | 31 | | |
| 2017-01-03 15:22 | 10.3 | 312 | -2.8 | 617 | 35 | | |
| 2017-01-05 13:00 | 10.3 | 347 | -2.8 | 536 | 69 | | |
| 2017-01-23 15:20 | 10.2 | 61 | -3.0 | 629 | 30 | | |
| 2017-01-26 15:26 | 10.1 | 312 | -3.0 | 623 | 25 | | |
| 2017-02-15 13:08 | 8.6 | 42 | -3.1 | 640 | 43 | | |
| 2017-02-18 15:59 | 8.3 | 318 | -3.1 | 638 | 32 | | |
| 2017-02-23 14:43 | 7.7 | 242 | -3.1 | 699 | 33 | | |
| 2017-03-15 13:13 | 4.6 | 343 | -3.1 | 713 | 41 | | |
| 2017-03-17 15:15 | 4.2 | 47 | -3.1 | 691 | 47 | | |
| 2017-03-22 14:54 | 3.3 | 343 | -3.1 | 745 | 56 | | |
| 2017-03-30 14:13 | 1.7 | 166 | -3.1 | 722 | 37 | | |
| 2017-03-31 11:28 | 1.5 | 347 | -3.1 | 758 | 53 | | |
| 2017-06-08 05:53 | 9.7 | 306 | -2.7 | 708 | 58 | | |
| 2017-08-03 04:40 | 9.8 | 165 | -2.6 | 665 | 65 | | |
| 2017-09-07 04:54 | 6.6 | 82 | -2.7 | 565 | 71 | | |
| 2017-12-16 15:35 | 6.8 | 342 | -3.1 | 538 | 69 | | |
| 2017-12-30 15:47 | 8.3 | 310 | -3.1 | 674 | 48 | | |
| 2018-01-03 16:00 | 8.6 | 45 | -3.2 | 690 | 28 | | |
| 2018-01-07 16:18 | 9.0 | 140 | -3.2 | 693 | 27 | | |
| 2018-01-10 15:51 | 9.2 | 26 | -3.2 | 735 | 92 | | |
| 2018-01-13 16:00 | 9.4 | 278 | -3.2 | 676 | 31 | | |
| **2018-01-18 16:05** | 9.7 | 216 | -3.2 | 827 | 45 | | |
| 2018-03-18 16:01 | 8.7 | 338 | -3.4 | 639 | 46 | | |
| 2018-03-22 16:13 | 8.2 | 75 | -3.4 | 656 | 43 | | |
| 2018-04-23 11:07 | 3.1 | 66 | -3.4 | 810 | 46 | | |
| 2018-04-24 10:05 | 2.9 | 261 | -3.4 | 770 | 34 | | |
| 2018-05-05 10:14 | 0.8 | 342 | -3.4 | 759 | 50 | | |
| **2018-05-10 10:19** | 0.4 | 281 | -3.4 | 1071 | 100 | | |
| 2018-05-12 06:49 | 0.7 | 298 | -3.4 | 776 | 53 | | |
| 2019-03-04 16:24 | 10.6 | 97 | -2.9 | 704 | 54 | | |
| 2019-03-18 16:30 | 10.7 | 66 | -2.8 | 711 | 45 | | |
| 2019-03-20 16:33 | 10.7 | 113 | -2.8 | 679 | 49 | | |
| 2019-03-21 15:19 | 10.7 | 306 | -2.8 | 684 | 55 | | |
| 2019-04-04 15:00 | 10.2 | 272 | -2.8 | 726 | 58 | | |
| 2019-04-05 15:57 | 10.1 | 123 | -2.8 | 711 | 48 | | |
| 2019-04-26 15:52 | 8.1 | 76 | -2.8 | 744 | 56 | | |
| 2019-05-14 11:26 | 5.4 | 102 | -2.9 | 732 | 93 | | |
| 2019-05-18 12:05 | 4.6 | 204 | -2.9 | 691 | 139 | | |
| 2019-05-20 09:44 | 4.3 | 231 | -2.9 | 676 | 156 | | |
| 2019-06-08 12:52 | 0.5 | 166 | -2.8 | 805 | 129 | | |
| 2019-07-08 06:28 | 5.4 | 97 | -2.8 | 734 | 90 | | |
| 2019-07-15 09:18 | 6.6 | 109 | -2.8 | 655 | 93 | | |
| 2019-08-26 05:30 | 10.8 | 342 | -2.6 | 667 | 70 | | |
| 2019-09-25 05:07 | 10.6 | 322 | -2.5 | 664 | 67 | | |
| 2019-10-29 03:11 | 8.0 | 18 | -2.4 | 748 | 138 | | |
| 2020-05-10 16:03 | 10.3 | 330 | -1.3 | 714 | 46 | | |
| 2020-05-21 15:50 | 9.4 | 47 | -1.3 | 728 | 20 | | |

| Date | | | | | | | |
|---|---|---|---|---|---|---|---|
| 2020-05-23 15:54 | 9.1 | 94 | -1.3 | 754 | 22 | | |
| 2020-05-25 15:27 | 8.9 | 138 | -1.3 | 741 | 22 | | |
| 2020-05-31 15:51 | 8.1 | 283 | -1.3 | 757 | 27 | | |
| 2020-06-13 15:30 | 6.1 | 46 | -1.3 | 747 | 26 | | |
| 2020-06-15 15:44 | 5.8 | 95 | -1.3 | 771 | 29 | | |
| 2020-06-17 16:05 | 5.4 | 145 | -1.3 | 765 | 26 | | |
| 2020-06-20 09:49 | 4.9 | 343 | -1.3 | 750 | 24 | | |
| 2020-06-27 11:46 | 3.5 | 342 | -1.3 | 842 | 102 | 440 | 104 |
| 2020-07-01 10:49 | 2.7 | 71 | -1.3 | 809 | 29 | | |
| 2020-08-02 10:05 | 4.0 | 101 | -1.3 | 677 | 37 | | |
| 2020-08-04 09:19 | 4.4 | 142 | -1.3 | 764 | 22 | | |
| 2020-08-12 09:14 | 5.8 | 329 | -1.3 | 731 | 22 | | |
| 2020-08-14 09:16 | 6.2 | 14 | -1.3 | 723 | 27 | | |
| 2020-08-15 10:05 | 6.4 | 227 | -1.3 | 730 | 26 | | |
| 2020-08-16 07:57 | 6.5 | 53 | -1.3 | 749 | 24 | | |
| 2020-08-18 08:18 | 6.9 | 103 | -1.3 | 762 | 22 | | |
| 2020-09-01 08:43 | 8.9 | 76 | -1.4 | 734 | 43 | | |
| 2020-10-03 05:53 | 11.2 | 81 | -1.3 | 734 | 25 | | |
| 2020-10-16 04:51 | 11.2 | 198 | -1.3 | 592 | 83 | | |
| 2020-11-03 04:57 | 10.5 | 259 | -1.2 | 767 | 27 | | |
| 2020-11-12 05:02 | 9.9 | 289 | -1.2 | 723 | 64 | | |
| 2021-04-28 15:54 | 10.8 | 129 | 0.4 | 765 | 26 | 313 | 125 |
| 2021-05-26 14:55 | 11.5 | 55 | 0.6 | 752 | 19 | 353 | 113 |
| 2021-05-27 15:10 | 11.5 | 261 | 0.6 | 832 | 83 | 516 | 109 |
| 2021-05-30 15:26 | 11.5 | 154 | 0.7 | 758 | 19 | 446 | 167 |
| 2021-06-01 15:39 | 11.4 | 203 | 0.7 | 745 | 19 | 351 | 162 |
| 2021-06-13 15:57 | 10.9 | 127 | 0.8 | 786 | 21 | 483 | 144 |
| 2021-07-25 09:39 | 5.4 | 343 | 0.9 | 921 | 66 | 556 | 162 |
| **2021-08-13 11:28** | 1.5 | 267 | 0.8 | 1048 | 32 | 724 | 164 |
| 2021-08-14 10:38 | 1.3 | 103 | 0.8 | 882 | 31 | 452 | 74 |
| 2021-08-19 11:24 | 0.3 | 47 | 0.8 | 967 | 29 | 483 | 88 |
| 2021-08-21 08:15 | 0.4 | 67 | 0.8 | 912 | 27 | 438 | 82 |
| 2021-08-26 09:33 | 1.4 | 14 | 0.8 | 913 | 98 | 537 | 96 |
| **2021-08-27 09:44** | 1.6 | 222 | 0.8 | 1166 | 39 | 923 | 100 |
| 2021-09-02 11:12 | 2.9 | 17 | 0.8 | 786 | 67 | | |
| 2021-09-22 10:15 | 6.8 | 120 | 0.7 | 799 | 24 | 452 | 80 |
| 2021-09-24 08:47 | 7.1 | 155 | 0.7 | 786 | 26 | 391 | 111 |
| 2021-11-24 05:07 | 11.3 | 295 | 0.6 | 751 | 24 | 466 | 88 |
| 2021-11-26 04:56 | 11.2 | 339 | 0.6 | 750 | 21 | 399 | 113 |
| 2022-05-17 15:51 | 9.7 | 115 | 2.1 | 780 | 106 | | |
| 2022-05-25 15:52 | 10.4 | 302 | 2.2 | 708 | 73 | | |
| 2022-06-10 15:45 | 11.4 | 315 | 2.3 | 738 | 73 | | |
| 2022-06-21 16:06 | 11.7 | 35 | 2.4 | 735 | 72 | 279 | 56 |
| 2022-06-23 16:05 | 11.8 | 80 | 2.4 | 768 | 65 | | |
| 2022-07-18 16:01 | 11.3 | 126 | 2.6 | 749 | 68 | | |
| 2022-07-24 15:42 | 10.9 | 264 | 2.6 | 719 | 82 | | |
| 2022-07-29 15:52 | 10.4 | 204 | 2.7 | 721 | 61 | | |
| 2022-07-31 12:29 | 10.3 | 222 | 2.7 | 718 | 90 | | |